\documentclass[twocolumn,showpacs,preprintnumbers,amsmath,amssymb]{revtex4}

\usepackage{graphicx}
\usepackage{dcolumn}
\usepackage{bm}

\begin{document}

\title{Slippage and boundary layer probed in an almost-ideal gas
by a nano-mechanical oscillator}
\author{M. Defoort, K.J. Lulla, T.Crozes, O. Maillet, O. Bourgeois, and E. Collin}

\address{
Universit\'e Grenoble Alpes, CNRS Institut N\'EEL, \\
 BP 166, 38042 Grenoble Cedex 9, France
}

\date{\today}

\begin{abstract}
We have measured the interaction between $^4$He gas at 4.2$~$K and a high-quality nano-electro-mechanical string device for its first 3 symmetric modes (resonating at 2.2$~$MHz,  6.7$~$MHz and 11$~$MHz with quality factor $Q > 0.1$ million) over almost 6 orders of magnitude in pressure. This fluid can be viewed as the best experimental implementation of an almost-ideal monoatomic and inert gas which properties are tabulated. The experiment ranges from high pressure where the flow is of laminar Stokes-type presenting slippage, down to very low pressures where the flow is molecular. 
In the molecular regime, when the mean-free-path is of the order of the distance between the suspended nano-mechanical probe and the bottom of the trench we resolve for the first time the signature of the boundary (Knudsen) layer onto the measured dissipation. Our results are discussed in the framework of the most recent theories investigating boundary effects in fluids (both analytic approaches and Monte-Carlo DSMC simulations).
\end{abstract}


\maketitle

%
%

Micro and nano-technologies have driven advances in various fields taking advantage of electronic, mechanical, and even nowadays {\it fluid} properties exploited at very small scales (micronic and sub-micronic) \cite{book}. The so-called area of micro and nano-fluidics is today under intense research, with applications ranging from chemistry to biology \cite{microfluidics,microfluidicsII,microfluidicsIII}.
Micro and nano-mechanical elements (MEMS and NEMS) are then key tools to probe and interact with gases and liquids, which are often simply air or water \cite{exampleSTM,otherExample,otherExampleII}. Especially with NEMS, the device non-invasivity can be pushed down to the sub-micrometer scale. 

Beyond technological applications and engineering problems, fundamental issues of fluid mechanics are also intimately associated to this research. These are essentially linked to the actual {\it interaction} between the fluid and a wall of some kind. In conventional, macroscopic and viscous fluid flow the boundary condition that is used to describe the physical phenomenon is the so-called no-slip property: at the level of the obstacle, the fluid is assumed to be clamped on irregularities of the surface and the tangential velocity goes to zero \cite{navier,flowbook}. This boundary condition becomes completely wrong in rarefied gases \cite{flowbook,flowbookII}, micro-nano fluidic devices \cite{microfluidicsIV,microChevrier} and quantum fluids such as mixtures of liquid $^3$He and $^4$He \cite{mixslip,mixslipII}.
A stunning and counter-intuitive example of this is water flow in carbon nanotubes demonstrating gigantic slippage \cite{science}.  

\vspace{0.2cm}
\begin{figure}
\includegraphics[height=5.7 cm]{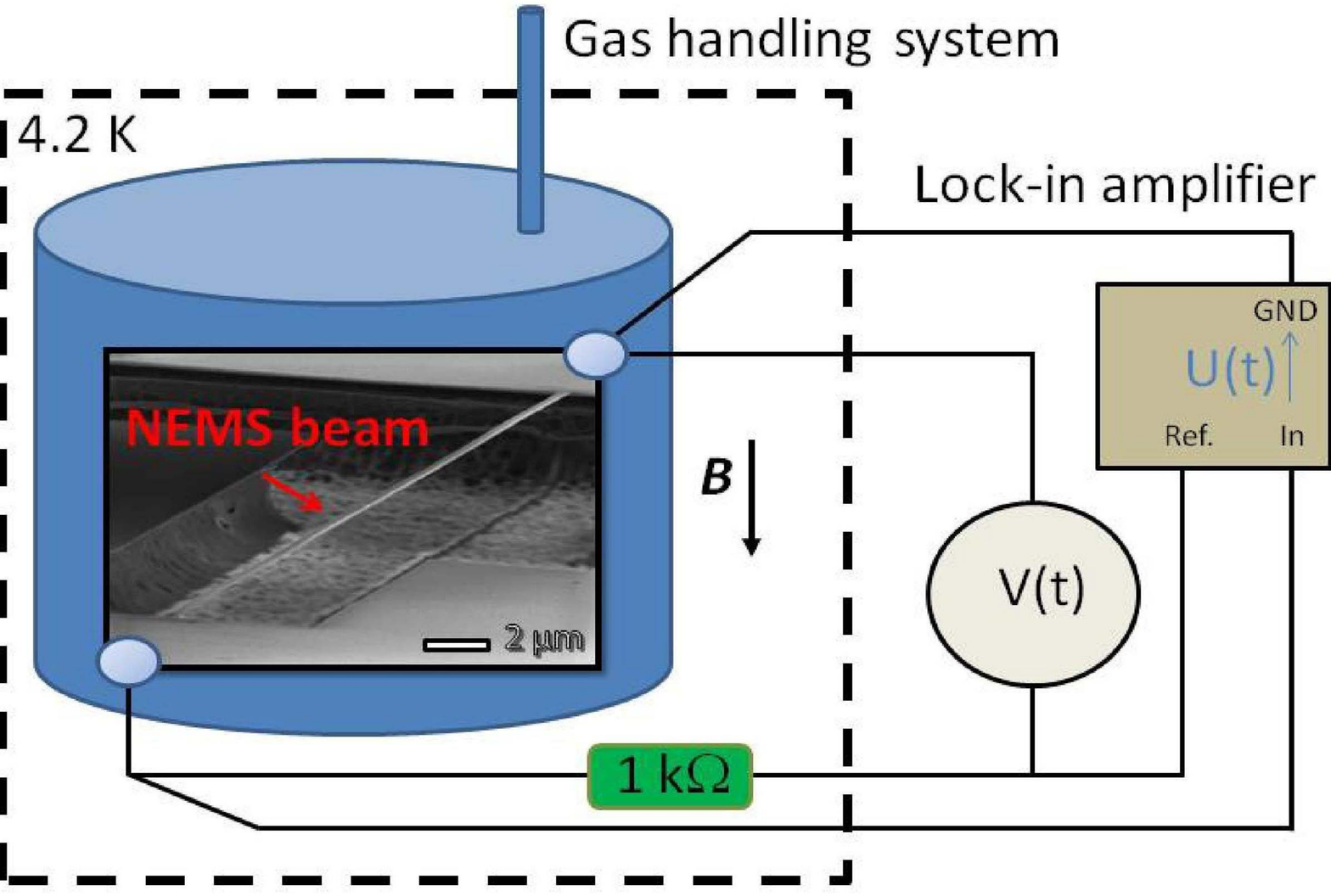}
\caption{\label{figure1} (Color online) Schematic of the experimental setup and SEM picture of the silicon-nitride nano-mechanical device. The dimensions are $w=300~$nm wide, $e=100~$nm thick and $h=100~\mu$m long. A thin layer of aluminum ($30~$nm) has been evaporated on top to create electrical contacts. The distance between suspended NEMS and bottom of the chip is $g=4~\mu$m. The whole cell is placed in a coil producing up to $1~$T magnetic fields $B$ in-plane with respect to the chip and perpendicular to the NEMS.}
\end{figure}

\vspace{1mm}
\baselineskip 12pt

Comprehending what happens physically between a fluid and a solid wall is thus essential for both practical applications and our fundamental understanding of fluid dynamics \cite{bullard,saderII}. With MEMS and NEMS this can be studied through oscillating flows in (more or less) confined geometries \cite{bullard,ekiI,ekiII}.
 In the boundary (or Knudsen) layer of thickness about one mean-free-path next to the solid surface, 
a rarefaction phenomenon occurs because particles collide more frequently with the wall than among themselves: this leads to a strong deviation of the statistical distribution of velocities from the Maxwellian equilibrium state (reached in the bulk). The flow is non-Newtonian, presents a nonzero tangential velocity (the slip effect) and reduced viscosity. There is today an extensive theoretical literature on the structure of the Knudsen layer, with even some predictions that {\it do not agree with each other} \cite{einzel,firstMD,theorKnusdenII,theorKnusdenIII,theorKnusden,lilley,saderII,slipagain}; at the same time direct experimental evidence of what actually happens within this boundary fluid was lacking. 

In this Letter we report on experiments conducted in $^4$He gas at 4.2$~$K using a nano-electro-mechanical device. 
We measure the friction experienced by the probe immersed in the fluid through one of its mechanical mode's resonance frequency-shift and broadening.
The three first symmetric modes have been used, and the pressure of the gas has been varied over almost six orders of magnitude.
In the rarefied gas limit, we measure for the first time a {\it decrease} of the damping acting onto the NEMS (with respect to the free molecular expression) which is evidence of the boundary layer effect. Two similar devices have been used with different distances from the suspended element to the bottom trench, thus proving unambiguously the boundary layer signature.  

%
%

\vspace{0.2cm}
\begin{figure}
\includegraphics[height=5.5 cm]{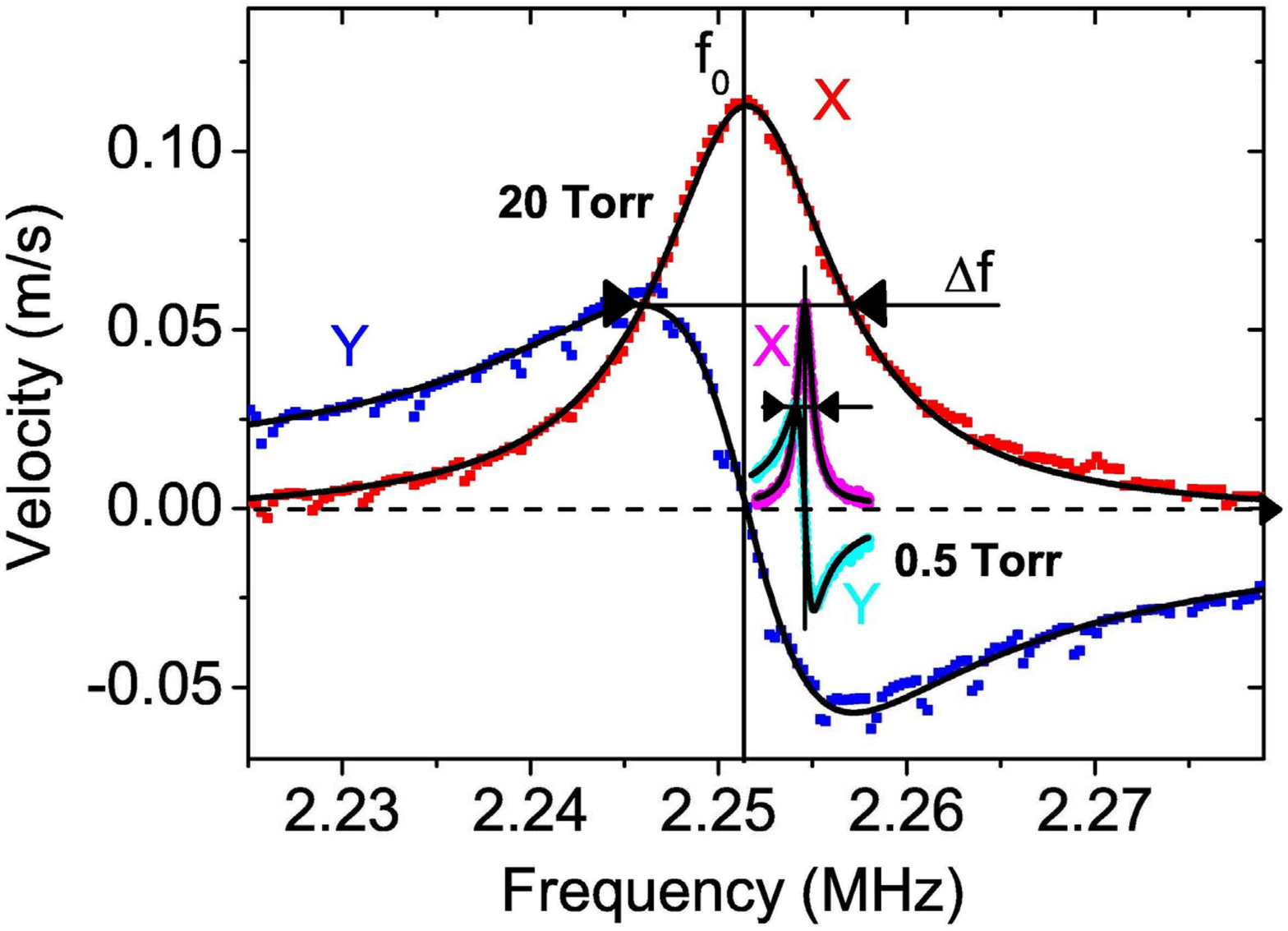}
\caption{\label{figure2} (Color online) Two typical resonance lines (mode \#1) obtained for pressures above and below the cross-over from Navier-Stokes to molecular flow (see Fig. \ref{figure3}). The lock-in amplifier leads to a homodyne detection giving access to the two quadratures $X$, and $Y$ \cite{rsi}. To keep the signal detectable, both magnetic field and current are increased as the damping increases (see text). Lines are Lorentzian fits giving access to resonance frequency $f_0$ and linewidth $\Delta f$.}
\end{figure}

\vspace{1mm}
\baselineskip 12pt

The NEMS fabrication and design can be found in Ref. \cite{kunalJLTP}. It is excited and detected by means of the magnetomotive scheme \cite{cleland}. 
A sinusoidal current $I_0 \cos(\omega t)=V(t)/1~\mbox{k}\Omega$ is fed through the suspended beam via a cold bias resistor, which produces a Laplace force thanks to a reasonably small magnetic field $B$. The out-of-plane motion is detected through the induced e.m.f. voltage $U(t)$ with a lock-in amplifier. 
A careful calibration procedure is used to deduce the applied forces and resulting velocities \cite{rsi}. 
The setup is schematically depicted in Fig. \ref{figure1}, and typical resonance lines are presented in Fig. \ref{figure2} for the first flexural mode. Each data point has been taken with at least two different excitation levels in order to make sure that heating and non-linear effects are negligible (see Supplemental Material \cite{supplement}). NEMS velocities $U_0$ have been kept small enough to ensure that the Reynolds number $R_e = \rho_g \, w \, U_0/\eta$ is smaller than 0.2 for all measurements.

When the mean-free-path of atoms or molecules is sufficiently short compared to experimental dimensions (object width, cavity width), the dynamics can be described by the well-known Navier-Stokes equation \cite{navier,PREpaper}. 
The first order correction to the dynamics when the mean-free-path becomes comparable to the size of the immersed objects is a modification of the boundary condition, introducing a phenomenological {\it slip length} \cite{einzel,mixslipII,theorKnusdenIII,lilley,saderII,slipagain}. 
In experiments this parameter is essentially a fit parameter (see e.g. \cite{science,microfluidicsIV,microChevrier,bouzigues,microChevrierII}), which theoretically depends on the nature of the surface interaction (diffusive or specular, see e.g. \cite{einzel,einzelJLTP}).

\vspace{0.2cm}
\begin{figure}
\includegraphics[height=6.5 cm]{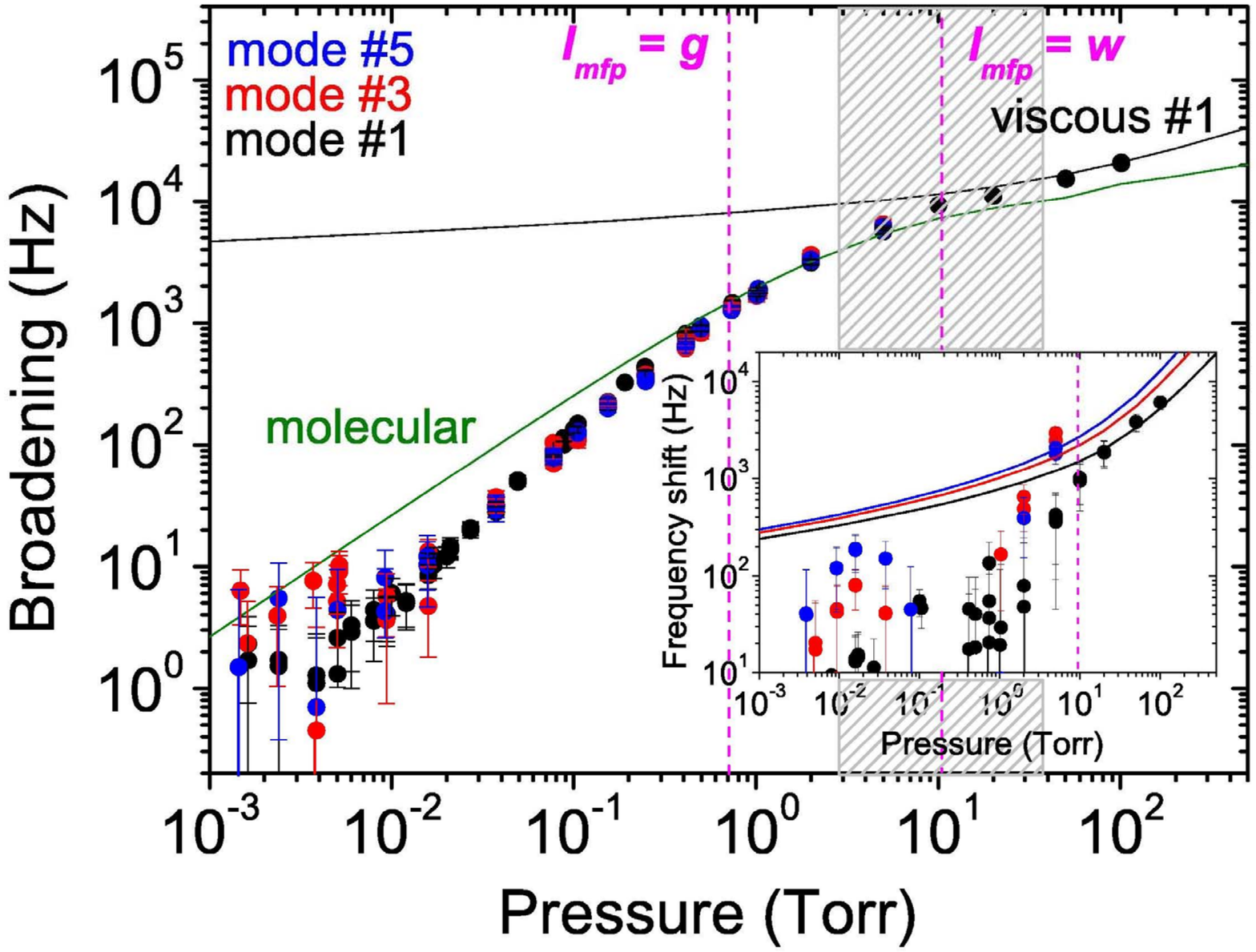}
\caption{\label{figure3} (Color online) Gas friction as measured by the NEMS device. The 3 colors stand for the first 3 symmetric modes (1, 3 and 5). Main: damping inferred from the broadening of the linewidth $\Delta f-\Delta f^{vac}$. The intrinsic contribution $\Delta f^{vac}$ has been subtracted. 
The black line is the calculation for the Navier-Stokes limit while the green (light) line corresponds to the molecular expression.  Inset: reactive contribution extracted from the frequency shift $f_0-f^{vac}_0$. Same conventions as main graph (plus Navier-Stokes calculations for modes \#3 and \#5). The dashed verticals mark the pressures when $l_{mfp} \approx g = 4~\mu$m and $l_{mfp} \approx w=300~$nm, and 
the shaded zone the region where the flow is neither Navier-Stokes nor molecular (see text). }
\end{figure}

\vspace{1mm}
\baselineskip 12pt

$^4$He gas at 4.2$~$K is the best experimental realization of a monoatomic and inert ideal gas. For pressures below a bar, the gas is classical (mean distance between atoms much larger than the thermal De Broglie wavelength). Thermodynamical properties including the scattering cross sections can be found in the literature \cite{nist,mit,vanderwaals,refs}.
Since the beam length $h$ is the largest lengthscale, we model all friction mechanisms as being local and integrate over the mechanical resonance mode shape to obtain the total damping parameter (see Supplemental Material for technical details \cite{supplement}). The interaction force per unit length between fluid and solid is written:
\begin{equation}
\frac{d F_{g}(\omega,z)}{dz} =+ \rho_g \, \omega^2 S_l \Lambda(\omega) \, \Psi_n(z) x_n(\omega) ,
\end{equation}
with $\rho_g$ the gas mass density, $S_l$ the characteristic cross-section presented by the beam and $\Psi_n(z) x_n(\omega)$ the displacement of the beam element at abscissa $z$ ($\Psi_n$ being the normalized mode shape, $x_n$ the overall harmonic motion amplitude for mode $n$ and $\omega = 2 \pi f$ the angular frequency). 
In the high-pressure gas, at low enough NEMS velocities the fluid dynamics is laminar and follows Navier-Stokes equation \cite{navier,eddyJLTP}.
Following Ref. \cite{saderBeam}, in the case of a rectangular beam we write for the damping coefficient $\Lambda(\omega)$:
\begin{equation}
\Lambda(\omega) = \Gamma(\omega) \, \Omega(\omega),
\end{equation}
where $\Gamma(\omega)$ is the well-known Stokes' (complex) function \cite{stokes} and $\Omega(\omega)$ a correction function valid for $e \ll w$ (beam thickness and width, respectively) \cite{saderBeam}.
Slippage is incorporated with the further renormalization \cite{mixslipII,hook}: 
\begin{equation}
\Lambda(\omega) \rightarrow 1+ \frac{1}{\left(\Lambda(\omega)-1\right)^{-1} - i \left( \frac{w/2}{\delta} \right)^2 \left( \frac{l_{slip}}{l_{slip}+w/2} \right) },
\end{equation}
with $\delta=2\sqrt{\frac{\eta}{\rho_g \, \omega}}$ the viscous penetration depth ($\eta$ being the dynamic viscosity of the gas) and $l_{slip}$ the slip length.
$l_{slip}$ can be expressed as a function of the specular fraction $\sigma$ of reflected particles from the probe's  surface  (1 is purely specular and 0 is diffusive) \cite{theorKnusdenII,theorKnusden,mixslipII,einzelJLTP}:
\begin{equation}
l_{slip} \approx 1.15 \left( \frac{1+\sigma}{1-\sigma} \right) \, l_{mfp}.
\end{equation}
The slip length is proportional to the mean-free-path and diverges for perfectly specular conditions on the immersed object.
We show in Fig. \ref{figure3} the calculation based on the model for mode \#1 with the reasonable value of $\sigma=0.5$. Above $l_{mfp} \approx w$ the curve fits the data rather well for both dissipative and reactive components (computed from imaginary and real parts of $\Lambda(\omega)$, respectively). For higher modes unfortunately, the damping was too large and no data could be acquired. Note that there is no other fitting parameter in the model (see Supplemental Material \cite{supplement}).

\vspace{0.2cm}
\begin{figure}
\includegraphics[height=6.7 cm]{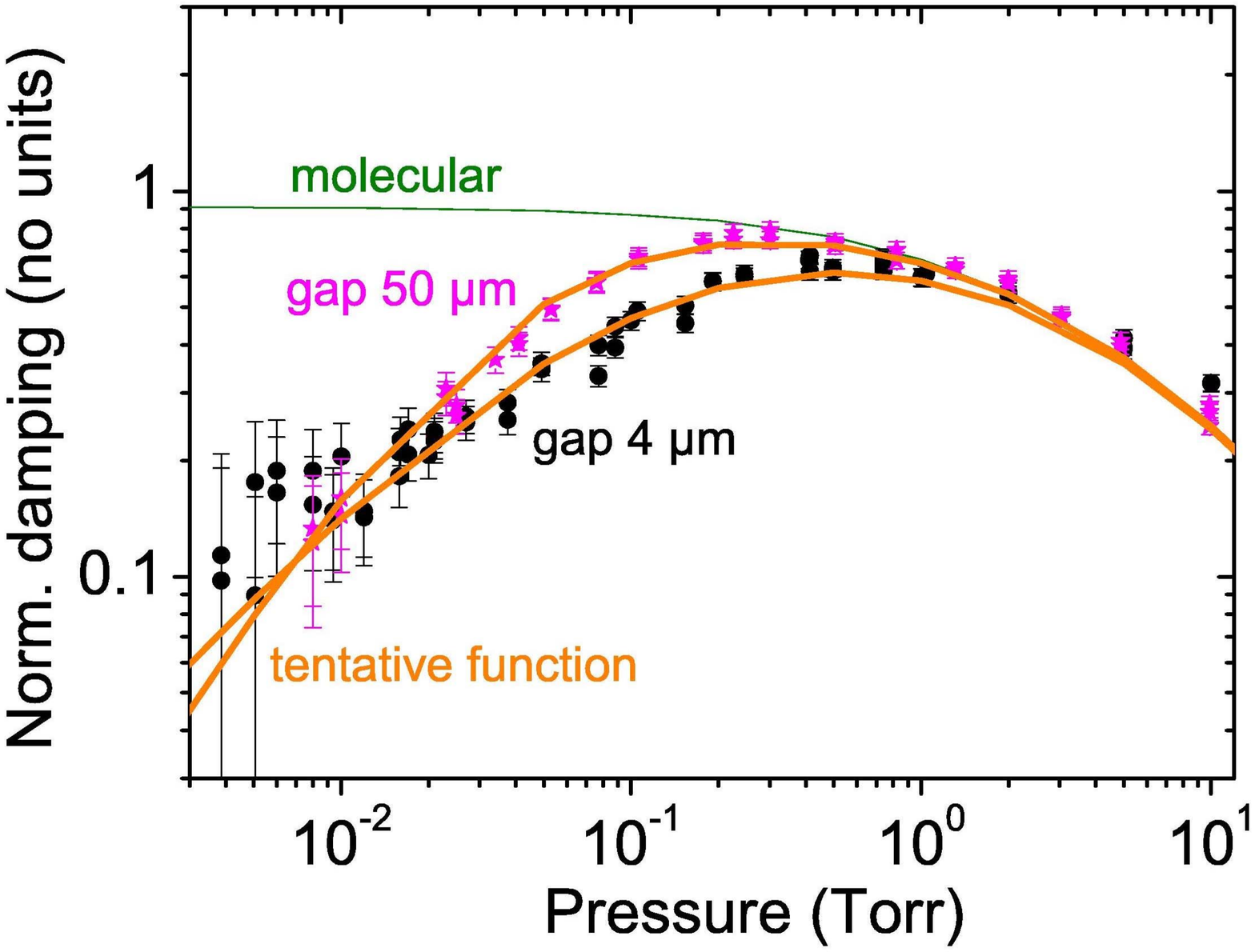}
\caption{\label{figure4} (Color online) Measured dissipation normalized to the ideal calculated molecular result of Eq. (\ref{equa1}). The first mode of the device of Fig. \ref{figure3} is shown (black dots) together with the data of another device having $g = 50\,\mu$m (magenta, or light stars). The green (light) line is the normalized green (light) line presented in Fig. \ref{figure3}. Below typically $l_{mfp} \approx g$ the measured friction {\it decreases} significantly. The two thick lines are tentative fits discussed in the text, with respectively $a=0.80 \pm 0.1$, $b=0.18 \pm 0.05$ ($g=4\,\mu$m) and $a=1.15 \pm 0.1$, $b=0.65 \pm 0.1$ (50$\,\mu$m). }
\end{figure}

\vspace{1mm}
\baselineskip 12pt

When the mean-free-path of atoms or molecules becomes very large compared to the transverse dimensions of the solid body in contact with the gas (container or obstacle), it is said to be in the {\it molecular flow} (or ballistic) limit \cite{flowbookII}. The same concept can be applied to a (classical or quantum) liquid, even though the elementary excitation will not be a bare atom or molecule, but a {\it quasi-particle} due to the strong interactions of each constitutive particle with its surroundings \cite{einzelJLTP}.
For a classical gas, few micro/nano fluidics experiments have been conducted in the molecular flow limit: usually the pressure range investigated corresponds to the slip flow described above, or the transition flow regime where neither Navier-Stokes nor molecular theories apply (shaded region in Fig. \ref{figure3}) \cite{microChevrier,microChevrierII,PREpaper}.   
However, in this pressure range an elegant similarity theorem has been recently demonstrated in Ref. \cite{bullard} describing the damping of the gas onto a nano-mechanical object.

In the molecular flow limit, the Navier-Stokes equation does not apply and the interaction between the moving body and the gas should be calculated from the transfer of momentum integrated over all particles bouncing off the device \cite{flowbookII, PREpaper,Yamamoto}. We write:
\begin{equation}
\frac{d F_{g}(\omega,z)}{dz} = + 2 \alpha \, \rho_g \, w \, \bar{v_g} \left[ i\, \omega \Psi_n(z) x_n(\omega) \right] , \label{equa1}
\end{equation}
with $\bar{v_g}$ the average thermal velocity in the gas and $\alpha$ a number close to 1 taking into account details of the scattering process (in particular $\sigma$, see Supplemental Material \cite{supplement}). $-i\, \omega \Psi_n(z) x_n(\omega)$ is the velocity of beam element $dz$, and the friction presents only a dissipative component. 
The calculation based on the cross-over expression given in Ref. \cite{PREpaper} is shown in Fig. \ref{figure3}. 
As expected from Eq. (\ref{equa1}), the measured damping is independent of frequency in the molecular regime, and the reactive component falls towards zero. 
Note that the frequency shift due to the mass of $^4$He adsorbed layers has been subtracted \cite{supplement}, which explains the rather large error bars in the Fig. \ref{figure3} inset.

Eq. (\ref{equa1}) predicts a damping proportional to pressure $P$ in the molecular limit. 
Note that within the simple modeling, the calculation has no free parameter \cite{supplement}. 
However, in Fig. \ref{figure3} the measured result is clearly {\it below} the calculation (up to an order of magnitude around 10$^{-3}~$Torr). 
The deviation occurs precisely for the pressure where the mean-free-path $l_{mfp}$ is of the order of the {\it distance between the NEMS device and the bottom of the chip} $g$, about  $ 4~\mu$m.
To quantitatively analyze this effect, we plot the normalized damping to the ideal molecular expression $2 \, \rho_g \, w \, \bar{v_g}$ in Fig. \ref{figure4}. At low pressures, the theoretical prediction tends towards the constant $\alpha$, which should be 1 for perfectly specular NEMS surfaces (green thin line).
To rule out any effects linked to the device itself, the measurements have been conducted at very low velocities and injected powers while the non-Maxwellian effects on Eq. (\ref{equa1}) due to the gas/NEMS boundary layer have been estimated: these are all negligible (see Supplemental Material \cite{supplement}).

The claim is thus that when $l_{mfp} > g$, the nano-mechanical probe gradually enters into the Knudsen layer attached to the bottom trench which {\it diminishes} the measured viscosity, as expected. In Fig. \ref{figure4} we also show the normalized data obtained for a similar device having $g = 50\, \mu$m. The decrease in the measured damping occurs clearly at lower pressures which validates the claim. To our knowledge, this effect has never been reported before while the intense recent theoretical investigations where clearly calling for experimental inputs (see e.g. \cite{saderII} and references therein).

Mathematically, the problem at hand is particularly tough and for most predictions it requires accurate (and demanding) numerical simulations. 
As a result, the most recent direct simulation Monte-Carlo methods (DSMC) \cite{saderII,lilley} contradict older findings: in particular the prediction of a reduction of the effective boundary layer viscosity to maximum 1/2 of its bulk value $\eta_{\infty}$ for a perfectly diffusive surface. 
Our experimental findings seem to contradict this point as well, since in Fig. \ref{figure4} the measured decrease of the effective viscosity rises up to a factor of 10 at the lowest pressures with no sign of saturation. In this sense, experiments are compatible with the most developed DSMC theories.

The quantitative analysis of the data can be pushed one step further, building again on the theoretical work of Refs. \cite{saderII,lilley}. The effective viscosity is supposed to scale as $\eta_{\infty}\,(g/l_{mfp})^{a}$ for large mean-free-paths. In order to match high and low pressure limits within the molecular range, we propose the phenomenological expression:
\begin{equation}
\mbox{damping} \propto P \, \frac{1}{1+b \, \left( l_{mfp}/g \right)^{a}}.
\end{equation}
This function is the one used on Fig. \ref{figure4} with $g$ equal to 4$\,\mu$m and 50$\,\mu$m for the two samples, with slightly different $a$ and $b$ fit parameters (see caption).
While the fits are quite convincing, the obtained $a$ exponent does not seem to match theoretical predictions, which calls for both new experimental and theoretical investigations.


In conclusion, we have measured the friction experienced by a nano-mechanical device immersed in an almost-ideal gas, $^4$He at 4.2$~$K. The pressure has been ranged from about $10^{-3}~$Torr where the flow is molecular, up to about 1 atmosphere where the gas is described by a laminar Navier-Stokes flow. The first 3 symmetric modes of the NEMS structure have been used to analyze the dependence to frequency/mode shape. We inferred that a rather large slippage occurs in the Navier-Stokes limit, consistent with a reasonably large fraction of specular reflections of particles off the probe. At low pressures in the rarefied gas limit, when the mean-free-path of atoms exceeds the distance to the bottom of the chip we measured a large deviation with respect to the ideal $\propto P$ damping molecular expression. We interpret the effect as the reduction of the effective viscosity occurring in the boundary layer, and try to consistently fit the data with respect to the most recent theoretical DSMC results. We indeed reproduce a power-law decrease of the effective viscosity at very low pressures (long mean-free-paths) which does not saturate to 1/2 of its bulk value. However, the decrease is much faster than the theoretically expected one. Our findings should help modeling the structure of the Knudsen boundary layer, which is extremely valuable as far as the comprehensive understanding of rarefied flows in micro and nano-systems is concerned.

The authors want to greatfully thank H. Godfrin for extremely valuable discussions, and J. Minet and C. Guttin for help in setting up the experiment. 
One of the authors (E.C.) also wants to acknowledge stimulating discussions with K. Ekinci and J.-P. Poizat.
We wish to thank T. Fournier for help in the microfabrication process. We acknowledge the support from MICROKELVIN, the EU FRP7 low temperature infrastructure grant 228464 and of the 2010 ANR French grant QNM n$^\circ$ 0404 01.

\end{document}